# BCS-like Superconductivity in MgCNi$_3$

J.-Y. Lin[1], P. L. Ho[2], H. L. Huang[2], P. H. Lin[1], Y.-L. Zhang[3], R.-C. Yu[3], C.-Q. Jin[3], and H. D. Yang[2]

[1]*Institute of Physics, National Chiao Tung University, Hsinchu 300, Taiwan ROC*

[2]*Department of Physics, National Sun Yat-Sen University, Kaohsiung 804, Taiwan ROC*

[3]*Institute of Physics, Center for Condensed Matter Physics and Beijing High Pressure Research center, Chinese Academy of Sciences, P. O. Box 603, Beijing 100080, PRC*

The low-temperature specific heat $C(T,H)$ of a new superconductor MgCNi$_3$ has been measured in detail. $\Delta C/\gamma_n T_c = 1.97$ is estimated from the anomaly at $T_c$. At low temperatures, the electronic contribution in the superconducting state follows $C_{es}/\gamma_n T_c \approx 7.96 \exp(-1.46 T_c/T)$. The magnetic field dependence of $\gamma(H)$ is found to be linear with respect to $H$. $T_c$ estimated from the McMillan formula agrees well with the observed value. All the specific heat data appear to be consistent with each other within the moderate-coupling BCS context. It is amazing that such a superconductor unstable to ferromagnetism behaves so conventionally. The Debye temperature $\Theta_D = 287$ K and the normal state $\gamma_n = 33.6$ mJ/mol K$^2$ are determined for the present sample.

PACS numbers: 74.25.Bt; 74.25.Jb; 74.60.Ec

The newly discovered superconductivity in MgCNi$_3$ has been a surprise [1]. Though with $T_c \leq 8$ K which is lower than that of the other new intermetallic superconductor MgB$_2$ [2], MgCNi$_3$ is interesting in many ways. Being a perovskite superconductor like Ba$_{1-x}$K$_x$BiO$_3$ and cuprate superconductors, MgCNi$_3$ is special in that it is neither an oxide nor does it contain any copper. Meanwhile, MgCNi$_3$ can be regarded as fcc Ni with only one quarter of Ni replaced by Mg and with C sitting on the octahedral sites. With the structure so similar to that of ferromagnetic Ni, the occurrence of superconductivity in MgCNi$_3$ is really surprising. Actually, there has been a theoretical prediction that MgCNi$_3$ is unstable to ferromagnetism upon doping with 12% Na or Li [3]. In this context, MgCNi$_3$ could be a superconductor near the ferromagnetic quantum critical point [4,5]. A possible magnetic coupling strength due to spin fluctuations was proposed [6]. Even more, a $p$-wave pairing in MgCNi$_3$ was suggested to be compatible with the strong ferromagnetic spin fluctuations [3]. If it were a $p$-wave superconductor, it would be the one with highest $T_c$ (e.g., compared to Sr$_2$RuO$_4$ with $T_c \leq 1.5$ K). To examine these interesting scenarios, fundamental properties have to be experimentally established. Nevertheless, there has been no reliable report on fundamental parameters like the Debye temperature $\Theta_D$. The values of coupling strength from different experiments were inconsistent with each others [7,8]. Nor does there exist a consensus on the superconducting pairing symmetry. NMR experiments revealed an $s$-wave pairing in MgCNi$_3$ [7], while the tunneling spectra indicated an unconventional pairing state [8]. In this paper, we present the detailed thermodynamic data and the derivations of some fundamental parameters from them. *It is found that MgCNi$_3$ possesses BCS-like C(T) in the superconducting state.*

The MgCNi$_3$ sample was prepared based on the procedure described in [1]. The starting materials were magnesium powder, glass carbon, and nickel fine powder. The raw materials were thoroughly mixed, then palletized and wrapped with Ta foil before sealed into an evacuated quartz tube. The sample was first sintered at 600°C for a short time and ground before further treated in a similar way at 900°C for 3 hours. The x-ray diffraction pattern revealed the nearly single phase of MgCNi$_3$ structure. Details of the sample preparation and characterization will be published elsewhere [9]. Temperature dependence of resistivity $\rho(T)$ showed a similar curve as reported in the literatures [1,10]. For the present sample, $\rho$=217 and 93 $\mu\Omega$ cm at $T$=300 and 10 K, respectively. It is well known that $T_c$ significantly depends on the real carbon content in the nominal MgCNi$_3$ [1,11]. Magnetization, specific heat, and resistivity measurements all showed a superconducting onset at about 7 K in the present sample. The resistivity transition width is 0.5 K, while thermodynamic $T_c$ determined from $C(T)$ is 6.4 K (see below). $C(T)$ was measured using a $^3$He thermal relaxation calorimeter from 0.6 to 10 K with magnetic fields $H$ up to 8 T. Detailed description of the measurements can be found in [12].

$C(T)$ of MgCNi$_3$ with $H$=0 to 8 T is shown in Fig. 1 as $C/T$ vs. $T^2$. The superconducting anomaly at $H$=0 is much sharper than that in Ref. [1], and clearly persists even with $H$ up to 8 T. It is noted that $C/T$ shows an upturn at very low temperatures. This upturn disappears in high $H$, which is a manifestation of the paramagnetic contribution like the Schottky anomaly. The normal state $C_n(T)=\gamma_n T + C_{lattice}(T)$ was extracted from $H$=8 T data between 4 and 10 K by $C(T, H$=8T$)=\gamma_n T + C_{lattice}(T) + nC_{Schottky}(g\sim H/k_B T)$, where the third term is a 2-level Schottky anomaly. $C_{lattice}(T)=sT^3+uT^5$ represents the phonon contribution. It is found that $\gamma_n$=33.6 mJ/mol K$^2$. This value of $\gamma_n$, with the electron-phonon coupling constant $\lambda$ estimated below, requires a higher band $N(E_F)$ than most of those reported from calculations [3,6,27,32]. $\Theta_D$ derived from $C_{lattice}$ is 287 K, impressively lower than that (450 K) of Ni. This low $\Theta_D$, nevertheless, is close to the estimate based on the softening of the Ni lattice [32], which could enhance the electron-phonon interaction. The concentration of paramagnetic centers can be estimated to be the order of 10$^{-3}$. With a dominant content of Ni in this compound, this number is understandable.

To elucidate superconductivity in MgCNi$_3$, it is of interest to derive $\Delta C(T) = C(T) - C_{lattice}(T) - \gamma_n T$. The resultant $\Delta C(T)/T$ at $H$=0 is shown in Fig. 2(a). By the conservation of entropy around the transition, the dimensionless specific jump at $T_c$ $\Delta C/\gamma_n T_c$=1.97±0.10 as shown in Fig. 2(b). This value of $\Delta C/\gamma_n T_c$ is very close to that in [1], though with a sharper transition in the present work. If the relation of $\Delta C/\gamma_n T_c=(1.43+0.942\lambda^2-0.195\lambda^3)$ [13] is adapted as was in Ref. [1], $\lambda$ is estimated to be 0.83. Both values of $\Delta C/\gamma_n T_c$ and $\lambda$ suggest that MgCNi$_3$ is a moderate-coupling superconductor rather than weak-coupling. To compare $\Delta C(T)$ of MgCNi$_3$ with a BCS one, $\Delta C(T)/T$ from the BCS model with $2\Delta/kT_c$=4 was plotted as the solid line in Fig. 2(a). There was no attempt to fit data with the BCS model. The choice of $2\Delta/kT_c$=4 instead of the weak-coupling value 3.53 was somewhat arbitrary and was to account for the larger $\Delta C/\gamma_n T_c$=1.97 than the weak-limit one 1.43. However, it is noted that already the data can be well described by the solid line, except the low temperature part of data which suffer contamination from the magnetic contribution. With this very magnetic contribution, it is difficult to check the thermodynamic consistency. Nevertheless, if the data below 3 K are replaced by the solid line, entropy is conserved as shown in the inset of Fig. 2(a). It is worth noting that $\Delta C(T)/T$ of MgCNi$_3$ is *qualitatively* different from that of Sr$_2$RuO$_4$, which is considered as a *p*-wave superconductor [33].

To further examine $C_{es} \equiv C(T,H) - C_{lattice}(T)$, $C_{es}(T)/\gamma_n T_c$ vs. $T_c/T$ for $H$=0 was plotted in Fig. 3. The fit of data between 2 and 4.5 K leads to $C_{es}/\gamma_n T_c$=7.96exp(-1.46$T_c/T$). Both the values of the prefactor and the coefficient in the exponent are typical for BCS superconductors. Since the magnetic contribution would make $C_{es}$ overestimated at low temperatures, the value of 1.46 in the exponent is probably slightly underestimated. This is in contrast to the case of MgB$_2$, in which $C_{es} \propto \exp(-0.38T_c/T)$ [12,14]. This small coefficient in the exponent for MgB$_2$ is usually attributed to a multi-gap order parameter.

In magnetic fields, $C_{es}(T,H) \approx C_{es}(T,H=0) + \gamma(H)T$ [15,16]. For a gapped superconductor, $\gamma(H)$ is expected to be proportional to $H$ [17]. For nodal superconductivity, $\gamma(H) \propto H^{1/2}$ is predicted [18]. Actually, $\gamma(H)$ of cuprate superconductors has been intensively studied in this context [19]. To try to figure out $\gamma(H)$ in MgCNi$_3$, $C(T,H)/T$ vs. $H$ at $T$=0.6 K and $\delta C(T,H)/T (\equiv C(T,H)/T - C(T,0)/T)$ vs. $H$ at 2 K is shown in Fig. 4(a) and (b), respectively. Data with $H \geq 4$ T are presented as the solid circles and shown in Fig 4(a). The data clearly follow a straight line passing through the origin, which suggests $\delta\gamma \propto H$. The magnetic contribution is rather significant for low field data at 0.6 K. The open circles represent data of $C/T$ corrected with the Schottky term estimated from the previously mentioned fitting. (The correction is negligible at high fields.) Apparently, the Schottky anomaly is only an approximation and can not totally account for the magnetic contribution at 0.6 K, especially for $H \leq 0.5$ T. At $T$=2 K, the magnetic contribution is not so significant as at 0.6 K. Thus $\delta C/T$ in all magnetic fields are shown as the solid circles. As seen in Fig. 4(b), all high field data can be well described by the straight line, indicating again a linear $H$ dependence of $\gamma$. Data below $H$=1 T begin to deviate from the linear behavior due to flux line interactions at low $H$ [15]. The straight line passes through the origin in Fig 4(a), which implies that the flux line interactions are relatively insignificant compared to the core contribution at very low temperatures. This trend was also observed in [15]. The slopes $d\gamma/dH$ in Fig. 4(a) and (b) are 3.17±0.02 and 3.15±0.08 mJ/mol K$^2$ T, respectively. These identical values at different temperatures suggest that the relation $\delta\gamma \propto H$ is genuine. Using $\gamma(H) = \gamma_n(H/H_{c2})$, $H_{c2}$=10.6 T for the present sample, which is close to that estimated from $dH_{c2}/dT_c$ determined by both ... and $C$ measurements according to the WHH formula [25]. This value is smaller than what was found in [10], probably due to different carbon contents since $T_c$ of the present sample is also lower than that in [10]. On the other hand, one could try to fit the data in Fig. 4(b) by $\delta\gamma(H) \propto H^{1/2}$. The results are represented by the dashed line in Fig. 4(b). Apparently, the data can not be well described in this manner, in contrast to the nice $\delta\gamma(H) \propto H^{1/2}$ relation found in cuprates [20-24]. A phenomenological fit of $\delta C/T(H) \propto H^n$ leads to $n=0.73$ (the dot line in Fig. 4(b)), similar to that in the dirty-limit $Y(Ni_{1-y}Pt_y)_2B_2C$ [34].

Due to the proximity of ferromagnetism, superconducting order parameter in MgCNi$_3$ was expected to be $p$-wave by [3] and others. However, it is noted that the $s$-wave superconductivity in weak ferromagnetism phase was once proposed [4]. Since there is no evidence for nodal lines of order parameter from the specific heat data, nature must have chosen the gapped order parameter like $x+iy$ if it was $p$-wave in MgCNi$_3$. To further investigate this issue, $T_c$ can be estimated by the McMillan formula

$T_c=(\hbar\tilde{\omega}_D/1.45)\exp\{-1.04(1+\lambda)/[\lambda-\mu^*(1+0.62\lambda)]\}$,

where $\mu^*$ characterizes the electron-electron repulsion [26]. Taking the Fermi energy $E_F \approx 6$ eV from the energy band calculations [3,6], $\mu^*$ is estimated to be 0.15, and $T_c=8.5$ K is estimated by the above McMillan formula with $\lambda=0.83$. This impressing agreement with the observed $T_c$ implies that the magnetic coupling strength $\lambda_{spin}$, if it existed, would be very small. This is consistent with the conclusion reported in [27]. For comparison, $\lambda_{spin}=0.1$ would probably lower $T_c$ to 3.7 K. Should such a small $\lambda_{spin}$ have turned the order parameter into $p$-wave pairing, the physics would have been unusual. Considering only the Ni $d$ contribution would effectively make $E_F$ smaller and thus lower $T_c$, leaving possible $\lambda_{spin}$ even smaller. ($E_F=4$ eV leads to $T_c=7.6$ K which is even closer to that of the present sample.) It is instructive to compare the physical parameters of MgCNi$_3$ with those of Nb$_{0.5}$Ti$_{0.5}$ and Nb, which are two s-wave superconductors. The results are listed in Table I. MgCNi$_3$ appears ordinary among these superconductors. $H_{c2}$ of Nb is much smaller than those of the others because Nb$_{0.5}$Ti$_{0.5}$ and MgCNi$_3$ are typical type II superconductors while Nb is nearly type II. (The coherence length $\xi \approx 5.6$ nm in the present MgCNi$_3$ sample, and the preliminary magnetization measurements suggest a penetration depth $\lambda_L=128-180$ nm [9].)

In conclusion, we have presented high quality data of $C(T,H)$ in MgCNi$_3$. Parameters like $\Delta C/\gamma_n T_c$, $\Theta_D$, and $\gamma_n$ are well determined. Both the analysis of the data themselves and the comparative studies with other $s$-wave superconductors show that all the specific heat data in MgCNi$_3$ are consistent with each other within the moderate-coupling BCS context. It is amazing that such a superconductor unstable to ferromagnetism behaves so conventionally.

We are grateful to B. Rosenstein for discussions on $p$-wave pairing. This work was supported by National Science Council, Taiwan, Republic of China under contract Nos. NSC91-2112-M-110-005 and NSC91-2112-M-009-046.

*Note added.* After submitting this paper, another recent paper appeared with the related issues [35]. The authors in Ref. [35] reached a similar conclusion of $s$-wave superconductivity in MgCNi$_3$ in the framework of the two band model.

**Table Captions**

TABLE I. Comparison between $MgCNi_3$, $Nb_{0.5}Ti_{0.5}$, and Nb. Parameters of $MgCNi_3$ are similar to those of $Nb_{0.5}Ti_{0.5}$ and Nb. Parameters of $MgCNi_3$ are from the present work, and those of Nb$_{0.5}$Ti$_{0.5}$ and Nb are from Refs. [26-29].

**Figure Captions**

FIG. 1.    $C(T,H)/T$ vs. $T^2$ of MgCNi$_3$ for $H$=0 to 8 T.

FIG. 2.    (a) $\Delta C(T)/T$ vs. $T$. The data are presented as the solid circles. The solid line is the BCS $\Delta C(T)/T$ with $2\Delta/kT_c$=4. Deviation at low temperatures from the solid line is due to the magnetic contribution of a small amount of the paramagnetic centers in the sample. Inset: entropy difference $\Delta S$ by integration of $\Delta C(T)/T$ according to the data above 3 K and the solid line below 3 K. (b) The dashed lines are determined by the conservation of entropy around the anomaly to estimate $\Delta C/T_c$ at $T_c$.

FIG. 3.    $C_{es}$ of MgCNi$_3$ in the superconducting state is plotted on a logarithmic scale vs. $T_c/T$. The straight line is the fit from 2 to 4.5 K.

FIG. 4.    Magnetic field dependence of (a) $C/T$ at $T$=0.6 K and (b) $\delta C/T$ at $T$=2 K. The straight lines are linear fits of the data for $H\geq 4$ T implying $\delta\gamma\propto H$. The open circles in (a) represent data of $C/T$ corrected with the Schottky term (see the text). In (b), the fitting range is from 1 to 8 T. Data below $H$=1 T deviate from the linear behavior due to flux line interactions at low $H$. The fits by $\delta\gamma(H)\propto H^{1/2}$ and by $\delta\gamma(H)\propto H^n$ are also shown as the dashed and the dot line respectively in (b) for comparison. The latter leads to $n$=0.73.

|  | MgCNi$_3$ | Nb$_{0.5}$Ti$_{0.5}$ | Nb |
| --- | --- | --- | --- |
| $T_c$ (K) | 6.4 | 9.3 | 9.2 |
| $\Delta C/\gamma_n T_c$ | 1.97 | ~1.9 | 1.87 |
| $\ln(\omega_D/T_c)$ | 3.79 | 3.23 | 3.40 |
| $2\Delta/kT_c$ | $\geq$4 | 3.9 | 3.80 |
| $H_{c2}$ (T) | 10.6 | 14.2 | ~0.2 |
| $\Theta_D$ (K) | 287 | 236 | 275 |
| $\gamma_n$ (mJ/mol K$^2$) | 33.6 (11.2/Ni) | 10.7 | 7.79 |

Table I.

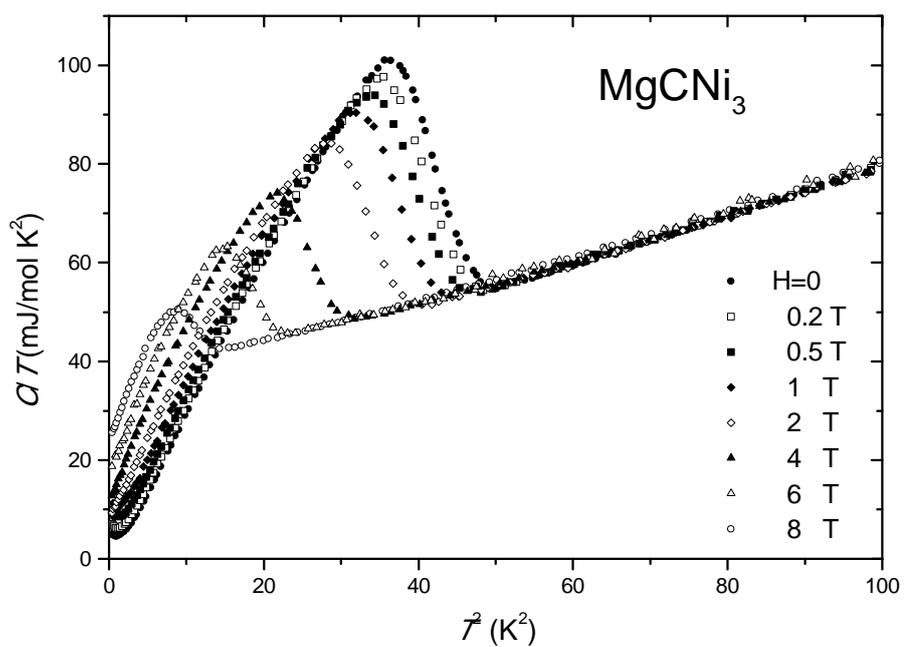

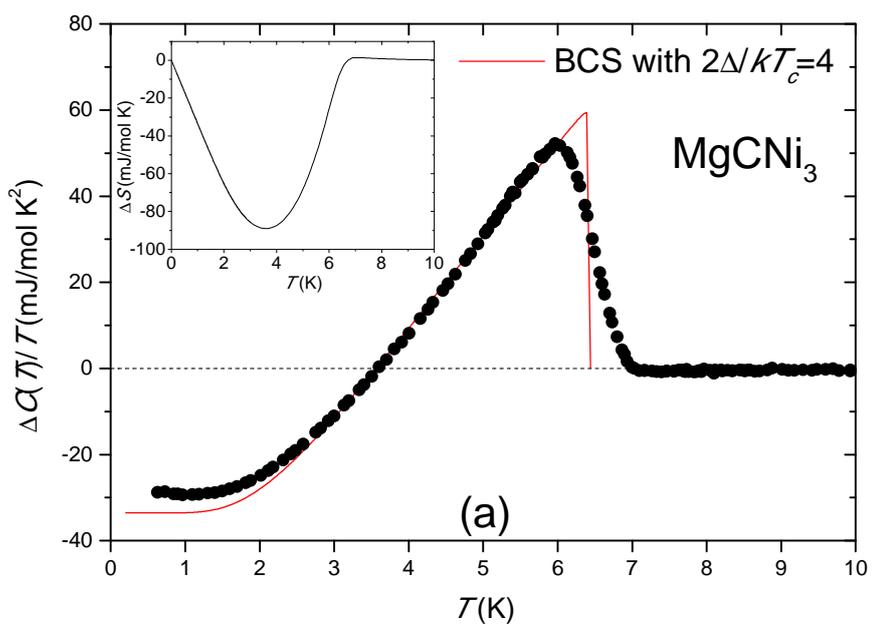

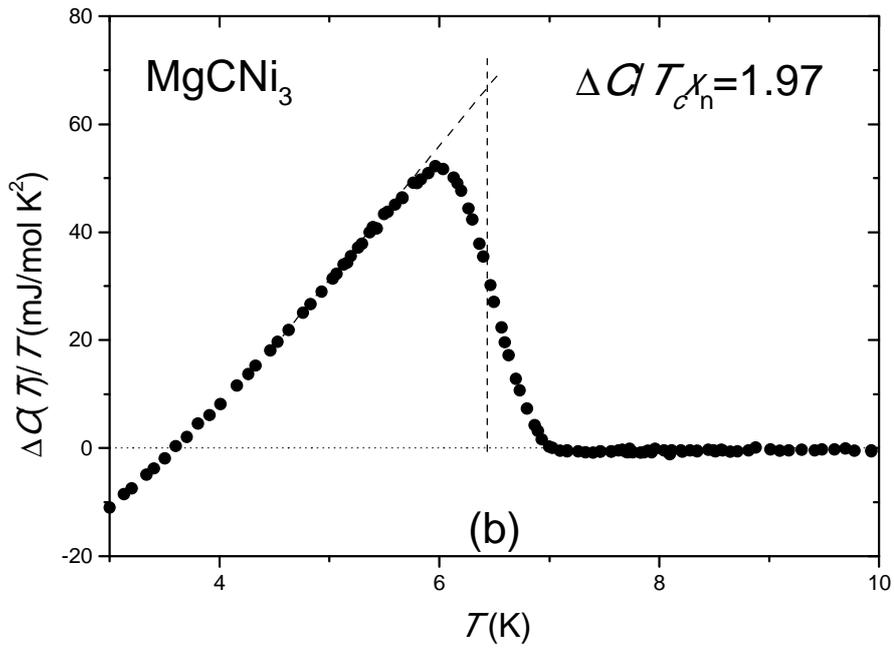

Lin et al. Fig. 2(b)

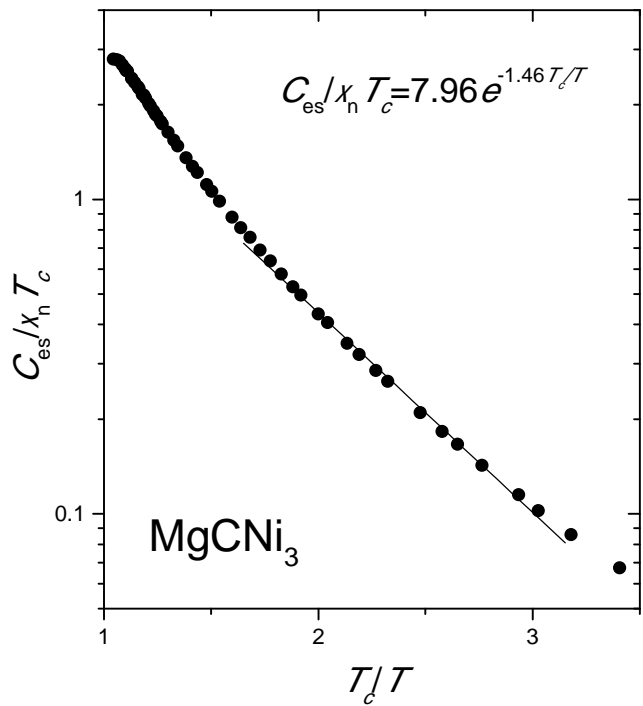

Lin et al. Fig. 3

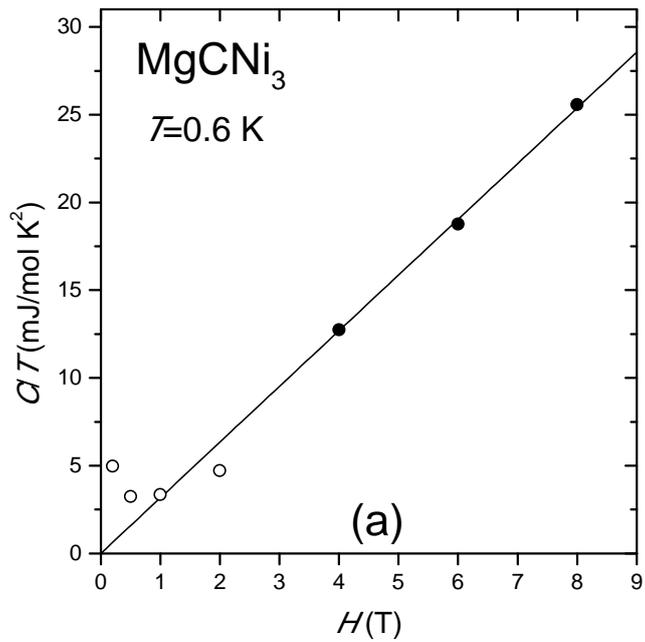

Lin et al. Fig. 4(a)

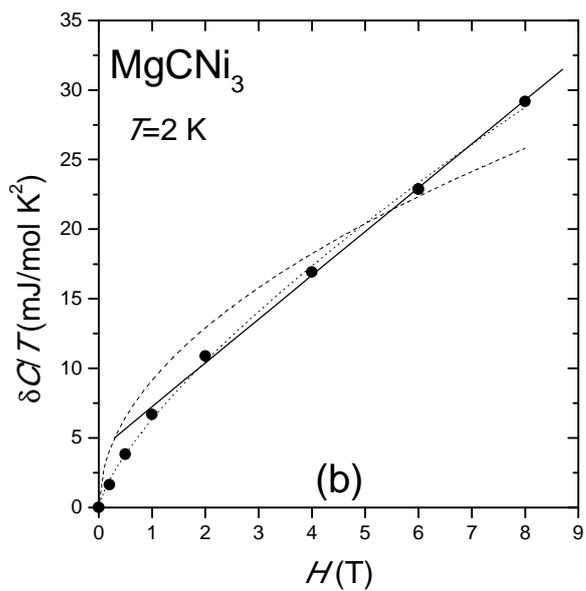

Lin et al. Fig. 4(b)

9